\documentclass[11pt]{article}
\usepackage[letterpaper,margin=1in]{geometry}
\usepackage{amsmath,amssymb}
\usepackage{graphicx}
\usepackage{times}
\usepackage[hidelinks]{hyperref}
\usepackage{caption}
\usepackage{placeins}
\graphicspath{{figs/}}
\title{Blind Field Validation of Passive Acoustic Vehicle-Speed Estimation
from the Convective Correlation Ridge}
\author{Anthony Raffaele \qquad Unnikrishna Pillai\\[2pt]
\small Department of Electrical and Computer Engineering,\\
\small Tandon School of Engineering, New York University, Brooklyn, NY 11201 USA}
\date{}

\begin{document}
\maketitle

\begin{abstract}
\noindent
Passive acoustic sensing is attractive for roadside traffic monitoring because it
is inexpensive, non-radiating, and simple to deploy. A microphone line beside a
road records two distinct delay structures. The direct sound wavefront arrives at
the speed of sound, and its inter-microphone delay at any instant is set by the
source position. Vehicle speed enters that delay only through its change over time.
The convective ridge follows the sound-energy pattern that travels along the array with
the vehicle, and its lag carries the speed directly. The gated
estimator reads speed only from the convective ridge and excludes the wavefront by
construction. A blind protocol fixes the estimator and its acceptance gates,
reduces a full recording to a frozen accepted set, and measures the independent
ground truth only afterward, with no calibration. On an hour of highway traffic
the accepted pass-bys with video agreed with that reference without a detectable
systematic bias. In a controlled low-speed experiment against
satellite positioning, the gated estimator evaluated at the pass times accepted at most
two of eleven passes, and these read below the logged speed by less than the stated accuracy of the satellite positioning. A simpler envelope read, chosen afterward, was broadly consistent with the
reference at a close offset and failed at a wider offset and higher speed, where
its lags fell far below the convective lag and no speed was recovered. The study is a proof of
concept for the measurement method rather than a finished product.
\end{abstract}

{\sloppy\noindent\textit{Keywords:} passive acoustic sensing; vehicle speed estimation;
microphone array; envelope cross-correlation; blind field validation; traffic
monitoring\par}

\section{Introduction}
Road noise is an inexpensive and non-radiating source of traffic information, and
roadside microphones are simple to deploy. Roadside recordings have been used for
traffic-state estimation \cite{Tyagi}, vehicle counting and direction
\cite{Zu}, ad-hoc microphone-array monitoring \cite{Toyoda}, and vehicle
observation from pass-by noise with a small microphone array \cite{Marmaroli}.
Acoustic vehicle-speed estimation is itself well developed. Two-microphone methods
read speed from the direct-sound delay \cite{LopezValcarce,Ishida}, and
single-point methods have been validated against video, passes at a controlled
speed, or lidar \cite{Cevher,Djukanovic,Kotus}. Speed from the cross-correlation
lag between sensors spaced along the road has also been used with pavement sensors
\cite{Zhang}. A field-validation method is still needed that establishes, without
calibration or retrospective tuning, that a roadside speed estimate is set by
vehicle motion rather than by sound propagation.

A linear roadside array records two distinct delay structures. The direct sound
wavefront produces an inter-microphone delay $\gamma_{mn} = (r_n - r_m)/c$,
where $r_m$ is the distance from the source to microphone $m$ and $c$ is the speed
of sound. For two microphones a distance $\Delta x$ apart this delay is bounded by
$|\gamma_{mn}| \le \Delta x / c$, and at each instant it is set by the source
position. Vehicle speed enters it only through its change over time, which
two-microphone methods exploit \cite{LopezValcarce,Ishida}. The convective ridge follows the vehicle-associated sound-energy
pattern that translates along the array, with lag $\tau = \Delta x / v$, where $v$
is the vehicle speed. The convective lag exceeds the wavefront bound by the factor
$c/v$ at every separation, so a larger aperture separates the two further in time.
An unconstrained correlator can lock onto short-lag structure, such as sound that
reaches all microphones almost together, so a speed estimate must isolate the
long-lag convective structure.

This paper contributes a field-validation method for acoustic speed estimation. The two
delay structures are separated explicitly, and the gated estimator reads speed only
from the convective ridge. The estimator and its acceptance gates are frozen and applied
blind, the recording is reduced to a frozen accepted set, and the independent
ground truth is measured only afterward, with no calibration. The method is then
assessed across two regimes: highway speeds against video, and low speeds at two
lateral offsets against satellite positioning. The paper reports where the method
held, where it accepted few passes, and where a simpler read failed.

\section{Array and convective estimator}
Four microphones recorded through one four-channel interface from a single
hardware clock, with no per-channel alignment. The highway recording used a
sampling rate of $48\,$kHz and lasted about one hour. The microphones sat on a
straight line parallel to the road at measured stations of $0$, $2.00$, $4.49$,
and $6.99\,$m.

\textit{The two ridges:} Let $a_m(t)$ be the energy envelope at microphone $m$,
which sits at station $x_m$. For straight motion parallel to the array at constant
speed, a pass-by feature attached to the source translates along the array as
$a_m(t) \simeq a_0(t - x_m/v)$, where $a_0(t)$ is the envelope at the station
origin. The cross-correlation of the envelopes at stations $x_m$ and $x_n$ then
peaks at
\begin{equation}
\tau_{mn} = \frac{x_n - x_m}{v}.
\label{eq:conv}
\end{equation}
The lag $\tau_{mn}$ differs from the wavefront delay, whose magnitude is bounded
by $\Delta x / c$. For constant speed parallel to the array the translation holds
at any lateral offset. Changes of speed, reflections, overlapping sources, and the
finite correlation window can displace the envelope-correlation peak.

\begin{figure}[t]
\centering
\includegraphics[width=0.6\linewidth]{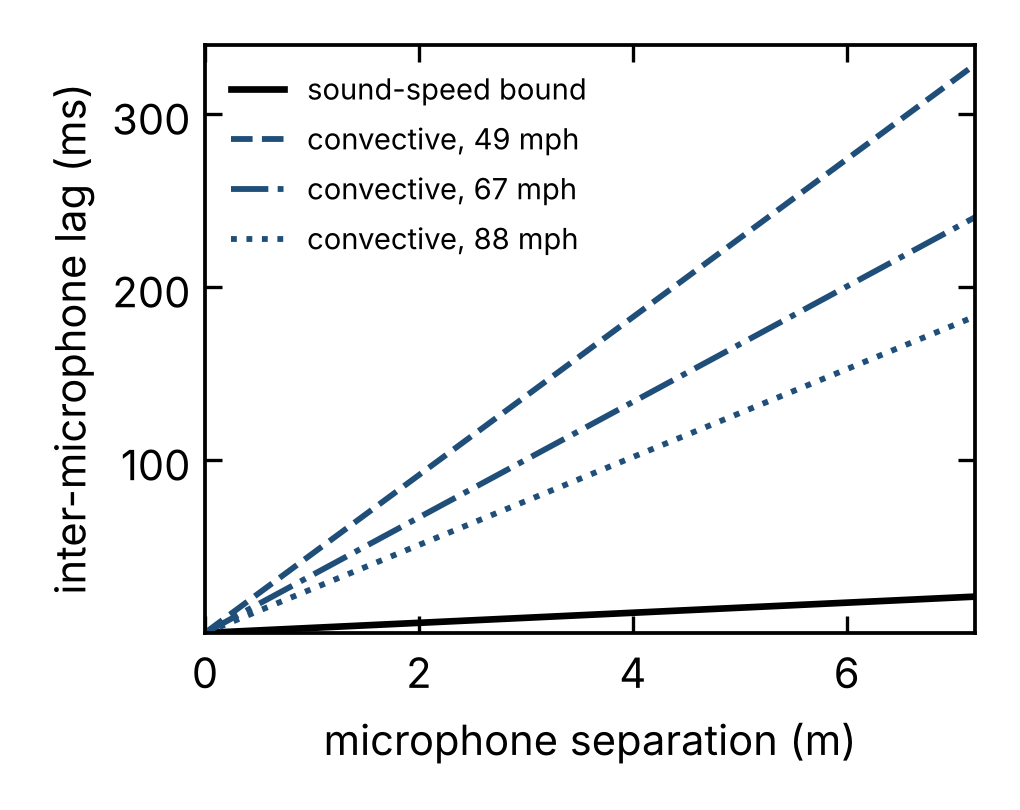}
\caption{Inter-microphone lag against microphone separation for the roadside
aperture. The solid black line is the sound-speed bound on the wavefront delay. The
three blue lines are the convective lag at the lowest, median, and highest
video-measured highway speeds. The convective lag is an order of magnitude larger, so
the estimator searches only the long-lag region.}
\label{fig:ridges}
\end{figure}

Figure~\ref{fig:ridges} compares the two delay structures for this aperture. On
the longest baseline the sound-speed bound is near $20\,$ms, whereas the
convective lag over the video-measured highway speeds is roughly $180$ to $320\,$ms. The
large separation lets the lag search exclude the wavefront entirely.

For detection, each channel is band-pass filtered to $400$--$2000\,$Hz, squared,
low-pass filtered at $10\,$Hz, and downsampled to $200\,$Hz, and the four results
are averaged. Pass-bys are found as peaks of this detection envelope, at a
prominence of at least $6\,$dB. A candidate needs a prominence of at least $8\,$dB, a width between $0.3$ and
$3\,$s at $3\,$dB below its peak, and isolation, meaning that no other detected
peak within $3\,$s is louder or less than $10\,$dB below it. Speed is then estimated from
the convective ridge with a fixed processing chain. Each channel is band-pass filtered to $50$--$400\,$Hz. This band was set in
advance from the low-frequency inter-channel coherence. It was kept after a
development check on one pass-by, made before any video was read, in which it
gave a far closer lag fit than the $400$--$2000\,$Hz detection band. The band-passed
signal is squared, low-pass
filtered at $10\,$Hz, and downsampled to $200\,$Hz to form an energy envelope.
Envelopes at each microphone pair are compared over a $3\,$s window centered on the peak of the four-microphone mean of these envelopes nearest
the detected pass-by, by zero-mean,
unit-energy normalized cross-correlation, the unweighted member of the
generalized cross-correlation family \cite{KnappCarter}. The lag search is
restricted to lags that correspond to vehicle speeds between $5$ and $60\,$m/s,
which excludes the wavefront neighborhood, so the wavefront delay cannot be
selected. Each pair lag is the highest interior local maximum in that range,
refined by parabolic interpolation. The six pair lags are fit against separation through the origin by least squares,
weighted by the correlation peak heights, and the fitted slope is the reciprocal of
the speed.

Acceptance uses six pre-specified quality gates, listed with their thresholds in
Table~\ref{tab:gates}. The through-origin fit must explain the six pair lags
closely. The six per-pair speeds must agree, measured by their coefficient of
variation. The mean speed from the three short baselines and the mean from the
three long baselines must agree, relative to the fitted speed. The fitted speed must
be a plausible road speed. The pass-by envelope peak must stand at least $6\,$dB above the
highest secondary peak of the detection envelope between $0.6$ and $3\,$s from it, counting only secondary maxima with a prominence of at least 3 dB. The speed must be stable when the
upper speed bound of the lag search is raised, a test called bound invariance,
which rejects a value set by the search boundary rather than by the data. The gates
and their thresholds were fixed before any ground truth was read.

\begin{table}[t]
\centering
\caption{The six acceptance gates and their thresholds.}
\label{tab:gates}
\begin{tabular}{cll}
\hline
Gate & Quantity & Threshold \\
\hline
1 & through-origin fit, centered coefficient of determination $R^2$ & $\ge 0.95$ \\
2 & per-pair speed, coefficient of variation & $\le 0.15$ \\
3 & short-versus-long baseline speed, relative difference & $\le 0.20$ \\
4 & fitted speed & $5$ to $60\,$m/s \\
5 & isolation depth on the detection envelope & $\ge 6\,$dB \\
6 & bound invariance, relative change in speed as the upper & $\le 0.05$ \\
  & bound rises from $60$ to $150\,$m/s & \\
\hline
\end{tabular}
\end{table}

\section{Blind protocol and ground truth}
For the highway validation, the protocol fixed the estimator and its thresholds
first, processed the full recording into a frozen accepted set second, and
measured the ground truth only afterward. After the accepted set was frozen, no
threshold, lag window, or event identity was changed in response to the reference.
For the highway reference, each vehicle was tracked across the video frame, and its
image rate was converted to road speed using the depth of its tire--road contact
point. That depth follows from the contact row, marked manually by an operator
blind to the acoustic speed, and from the camera geometry over a flat road. The
video reference carries a common-mode scale uncertainty of a few percent, with
camera height the largest contributor. In the controlled experiment a vehicle was driven past the array, and its speed
was logged by satellite positioning (GPS) at about one fix per second, with a
reported speed accuracy of a few meters per second. The reference for each pass is
the logged Doppler speed interpolated at the time the vehicle crossed the array,
and the log was synchronized with the audio to about $\pm 1\,$s. The controlled
passes were recorded at $48\,$kHz with four microphones on one four-channel
interface, at stations of $\pm 1.25$ and $\pm 3.50\,$m about the array center. The highway recording was made by the first author on the Wantagh State
Parkway in Nassau County, New York, and the controlled passes at a separate
low-speed site.

Writing $\hat v_k$ for the acoustic speed and $u_k$ for the reference speed of
pass-by $k$, the per-event fractional error is
\begin{equation}
\epsilon_k = \frac{\hat v_k - u_k}{u_k},
\label{eq:err}
\end{equation}
the median fractional bias is the median of $\epsilon_k$, and the mean absolute
error is the mean of $|\epsilon_k|$.

\section{Results}
\begin{figure}[t]
\centering
\includegraphics[width=0.52\linewidth]{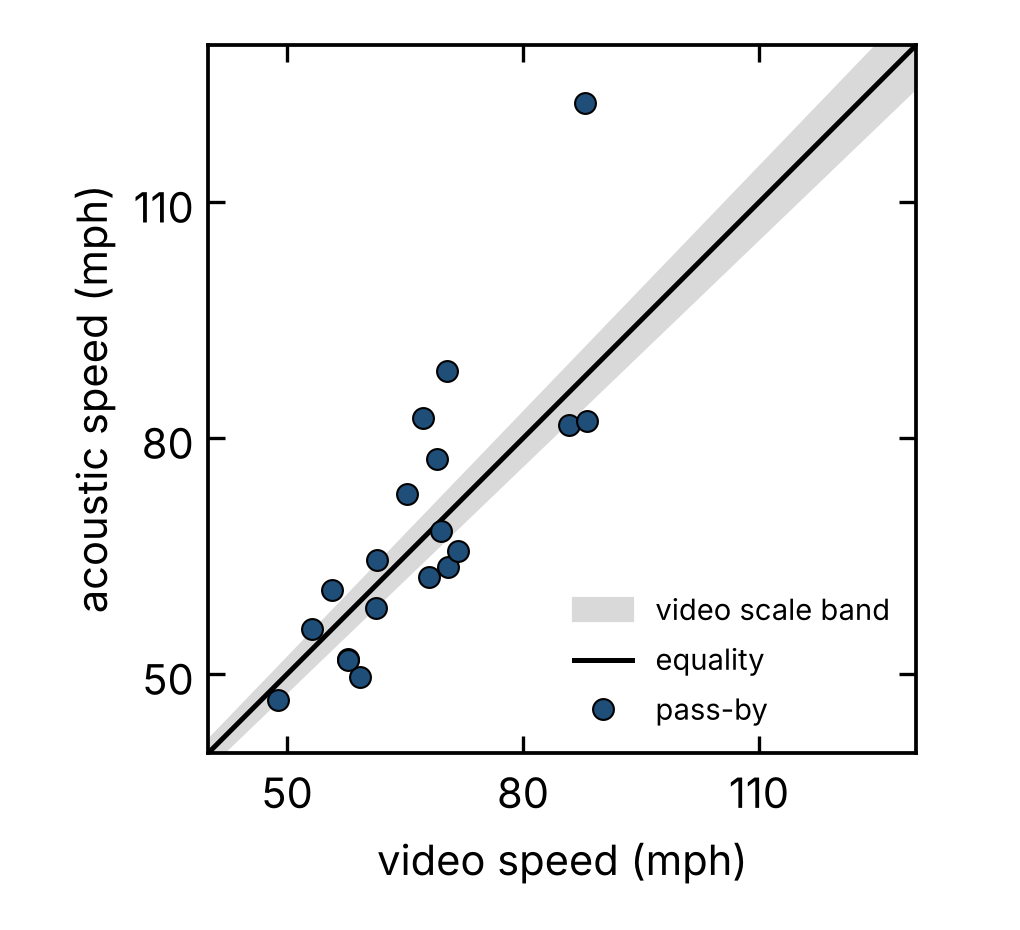}
\caption{Acoustic convective-ridge speed on the vertical axis against video ground
truth for the highway pass-bys. The solid line is equality, the shaded band is the
common-mode video scale uncertainty of about $\pm 4$ percent, and no calibration
was applied. The point
near the top is the retained pre-flagged pass-by.}
\label{fig:wantagh}
\end{figure}

Figure~\ref{fig:wantagh} shows the highway comparison. The detector found $305$
candidate pass-bys in the hour, and the frozen gates accepted $22$ of them. Most rejected candidates failed first at the fit gate,
gate~1. Two of the $22$ were
duplicate detections of the same envelope peak, and $19$ of the remaining $20$ had
usable video. Across these $19$ pass-bys the mean absolute error was about
eleven percent, the Pearson correlation with video was about $0.81$, and the
through-origin slope of acoustic speed on video speed was near unity. The median
bias was slightly negative, and its $95$ percent bootstrap interval contains zero,
so the data
do not establish a systematic offset. The video speeds ranged from about $49$ to
$88\,$mph. The point farthest from the equality line, near the top of
Figure~\ref{fig:wantagh}, is a pass-by flagged on acoustic grounds before the
ground truth was read and retained to preserve the frozen set. Its video
association later proved ambiguous, because several vehicles crossed within two
seconds of it. Without it the mean absolute error falls to about ten percent and
the slope to unity.

\begin{figure}[t]
\centering
\includegraphics[width=0.52\linewidth]{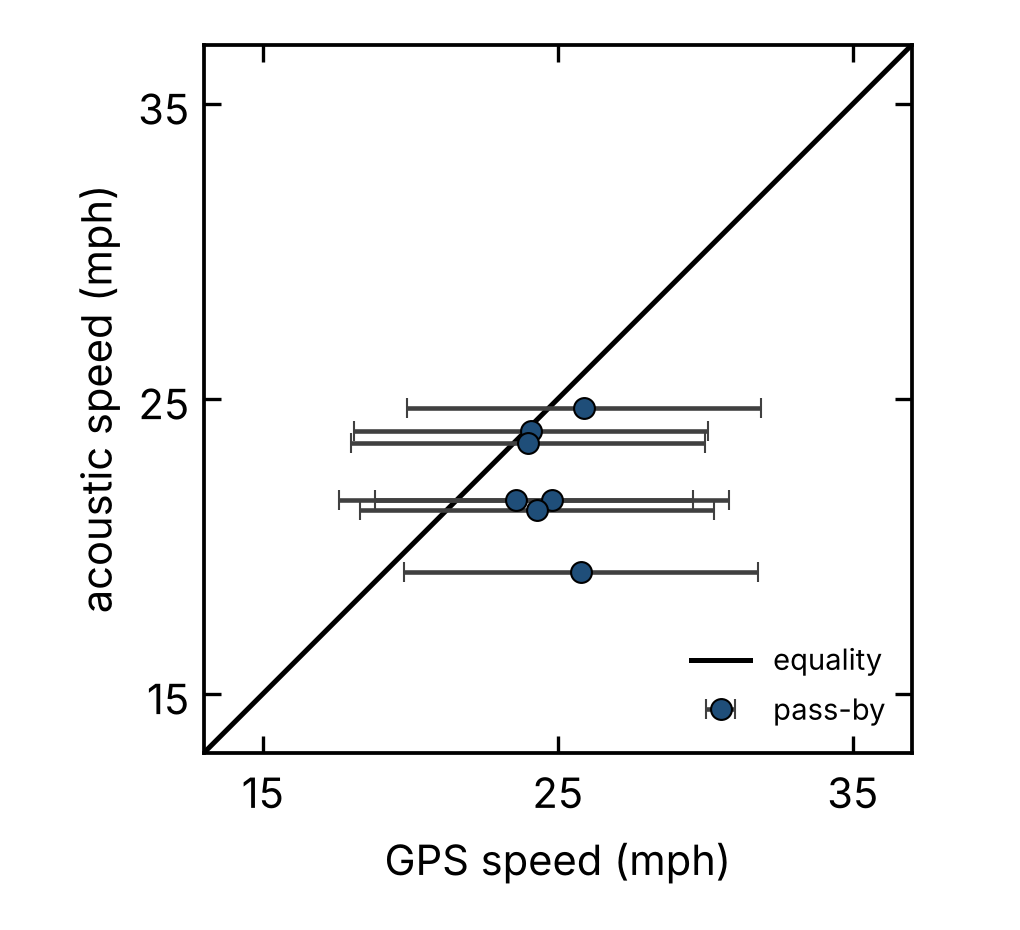}
\caption{Acoustic speed on the vertical axis against GPS ground truth for the
controlled low-speed passes at the $3\,$m offset. The solid line is equality, and
the horizontal error bars are the reported GPS speed accuracy of about
$\pm 6\,$mph. The acoustic speed was read with the simpler envelope read rather
than the gated highway method.}
\label{fig:controlled}
\end{figure}

Figure~\ref{fig:controlled} shows the controlled low-speed comparison. A vehicle
was driven past the array seven times near $25\,$mph at a lateral offset of
$3\,$m, with its speed logged by GPS. The controlled passes were analyzed at their
acoustic peak times, which were matched to the GPS crossings, rather than through
the highway detector. Evaluated there with the
same gates, the gated estimator rejected nine of the $11$ passes, five here and all
four at a wider offset of $6\,$m, on the fit and per-pair agreement gates. The other two
passes, both at the close offset, passed the five other gates. Their isolation
depths clear the $6\,$dB threshold whenever ripples shallower than about $1\,$dB
are ignored. Their speeds read $12$ and $15$ percent below the logged speeds, within
the GPS accuracy. This evaluation was
made after the logged speeds had been seen. Because the gated estimator accepted
at most two passes, the acoustic speeds in Figure~\ref{fig:controlled} come from a
simpler envelope read applied to all seven close passes. Each channel is band-pass filtered to $200$--$2000\,$Hz, squared, low-pass
filtered at $8\,$Hz, and downsampled to $200\,$Hz. Both filters are zero-phase
Butterworth filters. The six microphone pairs are compared by zero-mean,
unit-energy normalized cross-correlation over a five-second window centered on the
acoustic peak time of each pass, with no lag restriction and no gates. Separation is fit
against lag through the origin, with each pair weighted by its correlation peak
height. The read and its band were chosen after the gated estimator had been
evaluated on the passes and after the logged speeds had been seen, so the read is not blind. The
comparison is consistent with the convective observable and does not replicate the
gated estimator. The acoustic speed fell below the logged speed on all seven
passes, by about ten percent on average. The size of this shortfall depends on how
the read is built. Reversing the fit, as in the gated estimator, or forming the
envelope from the signal magnitude rather than its square reduces it, and one or
two passes then read above the logged speed. In Figure~\ref{fig:controlled} six of
the seven speeds lie within the GPS speed accuracy drawn as the horizontal error
bars, and one lies just outside it. The logged speeds span less than that accuracy,
so the comparison tests a single speed rather than tracking across speeds.

\begin{figure}[t]
\centering
\includegraphics[width=0.56\linewidth]{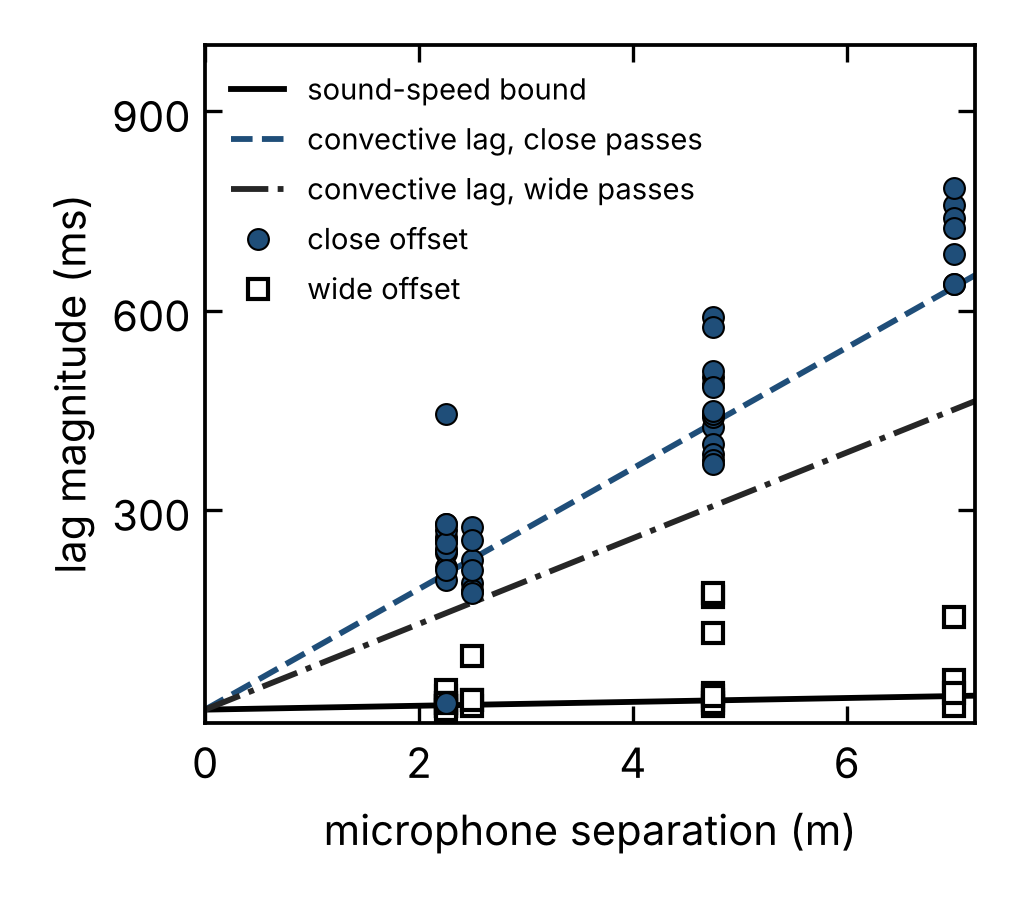}
\caption{Magnitude of the inter-microphone envelope lag against microphone
separation for all seven close-offset passes and all four wide-offset passes, read
with the simpler envelope read. The solid line is the sound-speed bound on the
wavefront delay. The dashed and dash-dotted lines are the convective lags predicted
by the mean logged speeds of the close and wide passes. Most close-offset lags lie
on or above their convective line. The wide-offset lags all fall below their convective line, most of them near
zero, and no speed is recovered from them.}
\label{fig:offset}
\end{figure}

Figure~\ref{fig:offset} shows where the simpler read fails. The four wide-offset
passes ran near $35\,$mph at a lateral offset of $6\,$m. Their lags all fall below
the convective lag that the logged speed predicts, most of them near zero, and they
no longer scale with separation, so the implied speeds far exceed the logged
speeds. Most close-offset lags lie on or above their convective line, in
keeping with the low readings in Figure~\ref{fig:controlled}. The convective lag in
Eq.~\eqref{eq:conv} does not depend on lateral offset, so the wider offset does not
by itself shorten it. The speed varied within these windows, but far too little to
shorten the lags this much. With no lag restriction, the read instead settled on
structure that reached all microphones almost together, and the source of that
structure was not identified. One wide pass began from a standstill about $6\,$s
before it reached the array, and another lies next to a gap in the GPS log.
Evaluated at the wide passes, the gated estimator also failed, with poor fits and
speeds far above the logged speeds, whereas at the close offset its fits
lay within about a quarter of the logged speeds. Because the wide passes differ
from the close passes in both offset and speed, no offset limit is claimed.

\textit{Limitations:} The reported highway accuracy describes the pass-bys that
passed the pre-specified gates, a small fraction of the candidates, and not all
traffic. The study covers one highway site over about one hour and one controlled
low-speed site, with most accepted pass-bys in the near lane. The GPS reference in
the controlled experiment is coarse relative to the acoustic errors, so it confirms
the acoustic speed only to within its own accuracy. The fine wavefront delay was
not measured reliably in this study, so the wavefront structure
serves only to justify excluding the direct sound from the speed search, not to
estimate range. No direction result is claimed from these data.

\FloatBarrier
\section{Conclusion}
A roadside microphone array can measure vehicle speed at highway speeds from the
convective correlation ridge for isolated pass-bys that produce a sharp envelope peak and
pass the pre-specified quality gates. The highway comparison used a blind,
uncalibrated protocol with the gated estimator, and the acoustic speed agreed with
video. The gated estimator accepted at most two of the $11$ controlled low-speed passes,
and these read below the logged speed within the GPS accuracy. A
controlled low-speed comparison, made with a simpler envelope read, gave
results broadly consistent with satellite positioning at a close offset, and the
same read failed at a wider offset and higher speed. Future work should add a second highway site, a lower-uncertainty speed reference
for the controlled regime, and a reliable direction measurement.

\section*{Acknowledgment}
AI tools were used mainly for editorial assistance including proofreading,
literature search and language refinement. All theoretical developments,
derivations and conclusions are the authors' work.

\end{document}